\documentclass[11pt,manuscript]{aastex}
\input psfig.sty

\newcommand {\ea} {{\it et~al.}}
\newcommand {\be} {\begin{equation}}
\newcommand {\ee} {\end{equation}}

\def\refitem{\par\parskip 0pt\noindent\hangindent 20pt}

\slugcomment{First draft: MS 08/13/00; Revisions: MB 08/21/00;
MS 08/31/00; RM 12/21/00; MB 1/27/01}
\shorttitle{Modeling the production of flares in $\gamma$-ray quasars}
\shortauthors{Sikora \ea}

\begin{document}

\title{Modeling the production of flares in $\gamma$-ray quasars}

\author{M.~Sikora and M.~B{\l }a\.zejowski}
\affil{Nicolaus Copernicus Astronomical Center, Bartycka 18, 00-716
Warsaw, Poland}
\email{sikora@camk.edu.pl}

\and

\author{Mitchell C.~Begelman\altaffilmark{1} and R.~Moderski\altaffilmark{2}}
\affil{JILA, University of Colorado, Boulder, CO 80309-0440, USA}

\altaffiltext{1}{also Department of Astrophysical and Planetary
Sciences, University of Colorado at Boulder}
\altaffiltext{2}{also Nicolaus Copernicus Astronomical Center, Warsaw}

\begin{abstract}

Theories of high energy radiation production in quasar jets can be
verified by studies of both time-averaged spectra and variability
patterns. While the former has been explored extensively, the latter
is in its infancy.  In this paper, we study the production of
short-term flares in the shock-in-jet model. We examine how the
flares' profiles depend on such parameters as shock/dissipation
lifetime, electron-injection time profile, adiabaticity, and
half-opening angle of the jet. In particular, we demonstrate the large
difference between flare profiles produced in the radiative and
adiabatic regimes. We apply our model to the $\sim$day timescale
flares observed in optically violently variable (OVV) quasars,
checking whether the external-radiation-Compton (ERC) model for
$\gamma$-ray flares at energies $> 30$ MeV (EGRET range) can be
reconciled with the flares observed at lower energies. Specifically,
we show that the strict correlation between X-ray and $\gamma$-ray
flares strongly supports the dominance of the synchrotron self-Compton
mechanism in the X-ray band. We also derive conditions that must be
satisfied by the ERC model in order to explain a lag of the
$\gamma$-ray peak behind the optical one, as claimed to be observed in
PKS 1406-076.  Finally, we predict that in ERC models where the MeV
peak is related to the break in electron distribution due to
inefficient cooling of electrons below a certain energy, the flares
should decay significantly more slowly in the soft $\gamma$-ray band than
at energies greater than $30$ MeV.

\end{abstract}


\section{INTRODUCTION}

The blazar phenomenon is well established to be related to nonthermal
processes taking place in relativistic jets on parsec/subparsec scales
and viewed by observers located within or nearby the Doppler cone of
the beamed radiation.  Thus, studies of the properties of blazar
radiation, such as spectrum shape and variability, provide exceptional
tools for exploring the deepest parts of extragalactic jets, their
structure, physics and origin.  However, in order to accomplish this,
one needs first to identify the dominant process responsible for the
production of $\gamma$-rays.

The $\gamma$-rays, with a luminosity peak located at photon energies
$> 1$ MeV, form the high energy spectral component and, together with
the synchrotron component peaking in the IR--X-ray range, constitute
the characteristic two-component spectrum of blazars (von Montigny et
al. 1995; Fossati et al. 1998).  The $\gamma$-ray fluxes show rapid and
high-amplitude variability, and in many OVV and HP (highly polarized)
quasars, at least during their high states, they reach luminosities
10-100 times larger than in the lower energy spectral bands.

There are several radiation mechanisms that can contribute to the
production of $\gamma$-rays in relativistic jets. One, suggested by
K\"onigl (1981), is the synchrotron self-Compton (SSC) process.
Others involve Comptonization of external radiation fields. The
candidate fields are: direct radiation from the disk (Dermer \&
Schlickeiser 1993); broad emission lines (BELs) and near-IR radiation
of hot dust (Sikora, Begelman \& Rees 1994; B{\l }a\.zejowski et
al. 2000 [hereafter, B2000]); rescattered central X-ray radiation
(Blandford \& Levinson 1995); and externally rescattered/reprocessed
synchrotron radiation of the jet (Ghisellini \& Madau
1996). Gamma-rays can also be produced by synchrotron radiation of an
extremely relativistic population of electrons/positrons, with the
maximum random Lorentz factor required to be at least $10^8-10^9$.
Such a population of electrons/positrons can be the product of a
synchrotron pair cascade, induced by $> 10^8$ GeV protons via the
photo-meson process (Mannheim and Biermann 1992). In some
circumstances, a certain contribution to the $\gamma$-ray band can
also be provided by synchrotron radiation of muons (Rachen \&
M\'esz\'aros 1998; Rachen 2000) and protons (Aharonian 2000; M\"ucke
\& Protheroe 2000).

Hereafter we focus on quasars only. In these objects the energy
density of the external diffuse radiation field is very high and, as
amplified by $\Gamma^2$ in the jet comoving frame, where $\Gamma$ is
the bulk Lorentz factor of radiating plasma, can easily dominate over
the energy density of magnetic fields carried by the jet and the
energy density of synchrotron radiation produced in the jet (see,
e.g., Sikora 1997).  Thus, the radiative output of quasars is very
likely to be dominated by $\gamma$-rays produced by Comptonization of
external diffuse radiation fields.  This so-called external radiation
Compton (ERC) model naturally explains the location of the
$\gamma$-ray peak/break, which in OVV and HP quasars is observed to be
in the 1-30 MeV range.  The MeV break, according to the one-zone ERC
model version proposed by Sikora et al. (1994), simply matches the
break in the energy distribution of electrons caused by their
inefficient cooling at lower energies.  The location of this break
requires the flares to be produced $\sim 0.1 - 1.0$ pc from the core,
a distance which corresponds nicely with the observed flare time
scales $\sim days$, provided that the formula $r \sim \Gamma^2 c
t_{fl}$ applies, where $r$ is the distance from the central engine and
$t_{fl}$ is the time scale of the flare.

The specific distance range of $\gamma$-ray production can be related
to the distance range over which the collision of two inhomogeneities
moving with different velocities is completed. According to this
scenario, the variability of blazars is modulated by the central
engine via instabilities in the innermost parts of an accretion disk
or magnetic eruptions in the corona (Sikora \& Madejski 2000). If
separations and lengths of inhomogeneities are of the same order, then
the number of reverse-forward shock pairs enclosed within a
dissipative zone is about $\Gamma^2$ and the number of shocks observed
at a given moment is of order 1 (Sikora et al. 1997). This can explain
high-amplitude fluctuations of blazar light curves, with occasional
flares exceeding the background blazar radiation by more than a factor
3.

In this paper, we use the shock-in-jet model to check whether
$\gamma$-ray flares produced by the ERC process can be reconciled with
those observed at lower energies (\S 3). In \S {3.1}, we show that the
similar rates at which $\gamma$-ray and X-ray flares decay, as
observed in 3C279 (Wehrle et al. 1998), require the production of
X-rays to be dominated by the SSC process, as was suggested by Inoue
and Takahara (1996) and by Kubo et al. (1998). In \S {3.2}, we study
the possible time-lags of flares produced at different frequencies and
discuss conditions required to reproduce the optical--$\gamma$-ray lag
claimed to be recorded in PKS 1406-076 (Wagner et al. 1995). Finally,
in \S {3.3} we make predictions regarding the variability of
$\gamma$-rays at energies $> 30$ MeV vs. variability in the soft
$\gamma$-ray band, around 1 MeV. These predictions are aimed at
verifying whether our interpretation of the MeV break in terms of
electron cooling effects is correct.  These studies are preceded by
analyses of how the flare profile depends on such model parameters as
shock lifetime, particle injection function, adiabaticity and jet
opening angle (\S 2). Our results are summarized in \S 4.

\section{FEATURES OF THE MODEL}

\subsection{Model assumptions}

To study flares in blazars we adopt the shock-in-jet scenario, in
which individual flares are produced by shocks formed due to velocity
irregularities in the beam and traveling down the jet with
relativistic speeds. Our model assumptions/approximations are 
(here, as elsewhere in the paper, primed quantities denote
measurements made in the co-moving frame of the source):

\refitem 
$-$ nonthermal plasma producing flares is enclosed within thin shells,
having a radial comoving width, $\lambda'$, much smaller than their
cross-sectional radius $a$;

\refitem
$-$ shells propagate down the conical jet with a constant Lorentz
factor $\Gamma$;

\refitem
$-$ magnetic fields, carried by the beam, scale with distance as $B'
\propto 1/r$;

\refitem
$-$ both magnetic field intensity and particle distribution are
uniform across the shell;

\refitem
$-$ relativistic electrons/positrons are injected into the shell
within a finite distance range, $\Delta r_{inj}$, which is equal to $c
\Delta t_{coll}$, where $\Delta t_{coll}$ is a timescale of the
collision;

\refitem 
$-$ the injection rate is parameterized by $Q = K \gamma^{-p}$, for
$\gamma_m < \gamma < \gamma_{max}$, and $Q \propto \gamma^{-1}$, for
$\gamma <\gamma_m$, where $\gamma$ is the random Lorentz factor of an
electron/positron;

\refitem 
$-$ radiative energy losses of relativistic electrons/positrons are
dominated by Comptonization of external radiation fields, synchrotron
radiation, and Comptonized synchrotron radiation;

\refitem
$-$ the observer is located at an angle $\theta_{obs} =1/\Gamma$ from
the jet axis.

The above assumptions can be justified as follows.  A shell in our
model approximates the geometrical site of relativistic electrons and
positrons. Such a site can be identified with the space between
forward and reverse shocks, formed due to the collision of two
inhomogeneities moving down the jet with different velocities. The
speed of the shell should then be identified with the speed of the
contact surface, while the width of the shell is just the distance
between forward and reverse shock fronts.  This distance is of course
a function of time and in the contact surface frame is given by the
formula $\lambda' = (\beta_f' + \beta_r') c \Delta t'$, where
$\beta_f' c$ and $\beta_r' c$ are forward and reverse shock
velocities, respectively, and $\Delta t'$ is the time since the
collision started. Comparing this with the cross-sectional radius of
the shell, $a=r/\Gamma$, one finds that
\be {\lambda'\over a} = (\beta_f' + \beta_r') {\Delta r_{inj} \over
r} . \ee
As was demonstrated by Komissarov and Falle (1997), for intrinsically
identical inhomogeneities (the same density and pressure) having
Lorentz factors differing by less than a factor two, the shock
velocities as measured in the contact frame are non-relativistic and,
therefore, our approximation $\lambda' \ll a$ is justified, provided
$\Delta r_{inj} \le r$.

Another geometrical assumption concerns the cone geometry. As VLBI
observations show, parsec scale radio jets have a tendency to be wider
at subparsec distances than on larger scales (Lobanov 1998) and,
within a finite distance range, they are much better approximated by
conical geometry than by cylindrical geometry. We should note that
even for $\Delta r_{inj}/r \ll 1$ it is important to take into account
the radial divergence of the jet, simply because radiation from shell
elements moving in different directions with respect to the observer
are differently Doppler boosted. (Of course, given possible departures
of the front and rear surfaces of the colliding inhomogeneities from a
spherical shape, one can expect that in general the shock surfaces
will not be exactly spherical. However, if these departures are
$\delta \lambda' \ll \Delta r_{inj}/\Gamma$, the effect on flare
profiles is negligible and the sphere approximation can be justified.)

In all of our models except those presented in Figure~\ref{flara}d the
half-opening angle of the jet is taken to be $\theta_j =
1/\Gamma$. This choice is somehow arbitrary, but our results are not
expected to be dramatically different if the real scaling is larger or
smaller by a factor three. There are several more or less direct
arguments that $\theta_j \sim 1/\Gamma$.  The deepest radio
observations of a quasar jet are those of Cygnus A, a radio galaxy that 
hides in its center a powerful quasar (Antonucci, Hurt, \& Kinney
1994). The jet opening angle in this object on 0.1 parsec scale is
about 7 degrees (Lobanov 1998; Krichbaum et al. 1998). This is equal
to the full Doppler angle, $2 /\Gamma$, if $\Gamma \sim 17$, whereas
bulk Lorentz factors of $\gamma$-ray quasars, as deduced from ERC
models, are enclosed within the range $10-20$ (see, e.g., Ghisellini
et al. 1998).  For $\theta_j \ll 1/\Gamma$, the SSC radiation,
calculated self-consistently assuming an ERC origin of $\gamma$-rays,
would exceed the observed soft X-rays; while $\theta_j \gg 1/\Gamma$
would increase jet energetic requirements up to a level difficult to
reconcile with theories of the central engine.  Furthermore, with
$\theta_j \gg 1/\Gamma$ the number of blazars would be too large
(Padovani \& Urry 1992; Maraschi \& Rovetti 1994).

We should also comment on our assumptions regarding the particle
acceleration process.  One is that the injection of relativistic
electrons is described by a two-power-law function, with a break at
$\gamma_m$. This break corresponds roughly with the average energy of
injected electrons, and the low-energy tail below the break mimics a
limited efficiency of the electron pre-heating process. It should be
noted that the model output parameters are insensitive to the exact
slope of this very hard low-energy tail, and that observationally it
can be imprinted only in the hard X-ray band. In the synchrotron
component it is invisible because of the synchrotron-self-absorption
process, and in soft--to--mid-energy X-rays it is obscured by the much
stronger SSC component.  Another assumption, which is not explicitly
listed above, concerns the acceleration time scale.  Since the time
scale of particle acceleration cannot be longer than the time scale of
radiative losses of even the most relativistic electrons, and since
the latter is known from modeling electromagnetic spectra of OVV/HP
quasars to be $\ll a/c$, the particle acceleration can be approximated
as an instantaneous process, represented in the electron kinetic
equation as a separate term called the injection function and denoted
by $Q$ (B2000). The particle acceleration time scale should not be
confused with the time scale of particle injection, a term which we
reserve hereafter for the time over which the shock operates, and
during which particles are accelerated continuously.

\subsection{Evolution of the electron energy distribution}

Evolution of electrons is described by a continuity equation (Moderski, Sikora \& Bulik
2000; B2000)
\be {\partial N_{\gamma} \over \partial r} = - {\partial \over \partial
\gamma} \left(N_{\gamma} {d\gamma \over dr}\right) + { Q\over c\beta \Gamma} ,
\ee
where the rate of electron/positron energy losses is
\be {d\gamma \over dr} = {1 \over \beta c \Gamma} \left(d\gamma \over
dt'\right)_{rad}- A{\gamma \over r} , \label{dgdr} \ee
$\beta = \sqrt {\Gamma^2 -1}/\Gamma$, and $dr = \beta c \Gamma dt'$.
The second term on the right-hand side of eq.~(\ref{dgdr}) represents
the adiabatic energy losses, with $A=1$ for 3D expansion and $A=2/3$
for 2D expansion.

In Figure~\ref{elevol} we show examples of electron evolution for the
case of radiation energy losses $(d\gamma /dt')_{rad} \propto
\gamma^2$.  Plotted is $N_{\gamma}\gamma^2 $ vs. $\gamma/\gamma_{c}$,
where $N_{\gamma}$ is the number of electrons per unit of energy and
$\gamma_{c}$ is the energy of electrons for which the radiative
cooling time scale is equal to the injection time scale. Noting that
the electron radiative energy loss time scales are given by
\be t_{rad} = \left \vert \gamma \over (d\gamma/dt') \right \vert_{rad}
\Gamma \ee
where $ \left \vert d\gamma/dt' \right \vert_{rad} = b \gamma^2$, one
can find that
\be \gamma_c = {\Gamma c \over b \Delta r_{inj}} . \ee
For $b$ dependent on $r$, $\gamma_c$ is computed for $b$ taken at 
$r = r_0 + \Delta r_{inj}$.

As one can see in Figure~\ref{elevol}, for $r <\Delta r_{inj}$ the
number of electrons is increasing, being saturated first at high
energies, then at lower energies. The slope of the saturated radiative
part of the electron energy distribution is given by $N_{\gamma} \sim
\gamma^{-p-1}$, while at energies $\gamma < \gamma_c$ the electron
distribution follows the injection function.  After injection stops,
the energy distribution of electrons above the break steepens and the
break itself moves down to $\sim \gamma_c$, or even lower energies,
depending on whether adiabatic losses are taken into account or not
(compare the left panels with the right panels).

We also illustrate in Figure~\ref{elevol} how the electron
evolution is different depending on whether the electron radiative energy losses
are constant or drop with distance.  In the latter case, the high
energy parts of the electron distribution saturate initially at lower
amplitudes, due to the fact that the energy losses are faster at the
beginning of the injection process than they are later.  This
amplitude increases with time, reaching a maximum at the end of the
injection process.

Finally, we note that the two cases considered here, one with
$d\gamma/dt' =$ const and one with $d\gamma/dt' \propto 1/r^2$, can be
related to two possible configurations for the source of radiation
surrounding the jet.  In the first case the radiation energy density
in the comoving frame of the shell is dominated by external radiation
sources (BEL-clouds or hot dust) located at a larger distance from the
central engine than the shell, while in the second case both are at a
similar distance and the energy density is scaled by $\xi L/r^2$, with
the fraction $\xi$ of the reprocessed/rescattered central luminosity,
$L$, being constant.

\subsection{Flares}

Using the algorithm developed by B2000 to compute the observed
radiation from relativistically and ``conically" propagating shells,
we analyze in this section the dependence of flare profiles on:
electron injection distance range, $\Delta r_{inj}/r_0$; injection
time profile, $Q(t')$; adiabaticity, $A$; and jet opening angle,
$\theta_j$.  Flares, presented in Figure~\ref{flara}, are produced by
the ERC process and are computed under the assumption that electron
radiative losses are strongly dominated by just this process. All
panels except (b) show flares which are produced in the radiative
regime, i.e., at an observed frequency that is contributed by
electrons with $\gamma \gg \gamma_c$.  One can easily check that
``radiative" flares reach maximum at a time when the observer is
receiving the first photon from a radius $r_0 + \Delta r_{inj}$. For
$\Delta r_{inj}/r_0 =1$, this is at $t/t_0 = 1$, where $t_0 =
r_0/(1+\beta)\beta c \Gamma^2 \simeq r_0/2 c \Gamma^2$.  The small
shifts of the peaks to the right of this value are caused by the
finite time of electron cooling.

As one can see from panel (a), flares produced by long lasting
injections, i.e., with $\Delta r_{inj} > r_0$, have very shallow
peaks. Such flares are never observed in blazars, suggesting that
$\Delta r_{inj} \le r_0$. On the other hand, those with very small
$\Delta r_{inj}/r_0$ are predicted to be more asymmetric than observed
flares, showing much faster increase than decay, the latter determined
by the light travel time effect related to the transverse size of the
shell.

Panel (b) shows ``adiabatic" flares, produced by electrons with
$\gamma \ll \gamma_c$.  One can see that even those with maximal
adiabatic losses (case $A=1$) are decaying significantly slower than
radiative flares.  This feature can be used to verify X-ray production
mechanisms (see \S {3.1}).

As panel (c) shows, the dependence of flare profiles on the time
profile of electron injection, $Q(t')$, is rather weak.  This is
because this dependence is diluted by the transverse-size light travel
time effect.  Only for $\Delta r_{inj}/r_0 > 1$, for very narrow jets
(see panel [d]) or for the observer located far from the jet axis can one 
try to recover the injection history $Q(t')$ from the observed light
curves. To better illustrate this effect we present in
Figure~\ref{flarb} flares produced by a double-triangle injection
profile. As one can see, the double-peak structure of the flare becomes
more diluted as the injection time becomes smaller or the observer moves out
from the jet axis.

\subsection{Frequency-dependent lags}

Very useful constraints on jet structure and radiation models can be
provided in the future by detailed observations of time lags between
flares detected in different spectral bands.  In this section, we
demonstrate the predictions of our model regarding the lags: at
different frequencies within the synchrotron component; at different
frequencies within the ERC component; and between synchrotron, ERC and
SSC components.

Our model parameters are chosen to match fiducial spectra of
$\gamma$-ray quasars during their high states (see, e.g., Fossati et
al. 1998). The spectrum evolution and lags predicted by our model are
presented in Figure~\ref{lags}. The observed frequencies of flare
production in the synchrotron component (panel a) and in the ERC
component (panel b) were chosen to be contributed by the same electron
energies at the beginning of the flare. At later moments this is no
longer true. This is because the synchrotron critical frequency is
proportional to the magnetic field intensity, $\nu_{syn} \propto
\Gamma \gamma^2 B'$, thus the synchrotron spectrum produced by
electrons from a given range of the energy distribution moves to the
left. In the ERC case, where $\nu_{ERC} \sim \Gamma^2 \gamma^2
\nu_{ext}$ and $\nu_{ext} \sim$ const ($\sim 10$ eV for BELs and $\sim
0.3$ eV for near IR radiation of hot dust), the spectrum doesn't drift
with time.  This is the main reason why synchrotron flares observed at
different frequencies are predicted to show lags (panel a), while the
effect is negligible in the ERC component (panel b): because the
critical synchrotron frequency is dropping, it takes longer to build
up the population of electrons necessary to produce a peak at higher
frequencies. Of course, one can see that the ERC flare profiles also
depend on frequency, but this is caused by the energy dependence of
the electron cooling rate, and the effect is significant only at
frequencies at which radiation is produced by electrons with $\gamma
\le \gamma_c$.

There are two additional factors that differentiate the profiles of
synchrotron flares from those of ERC flares. The first is that the
synchrotron cooling rate is $(d\gamma/dt')_{syn} \propto {B'}^2
\propto 1/r^2$, whereas $(d\gamma/dt')_{ERC}$ is assumed in this model
to be constant.  Since $r$ is increasing with time as the shock
propagates, this makes the synchrotron flare peak earlier. The second
is that the synchrotron radiation in the ``comoving" frame is
isotropic, whereas the ERC radiation is anisotropic (see Dermer 1995;
Sikora 1997; B2000). Direct comparison of flares produced by different
processes is provided in panel (d), at frequencies which are marked by
arrows in panel (c).

\section{CONFRONTATION WITH OBSERVATIONS}

\subsection{X-rays vs. $\gamma$-rays}

Unfortunately there were only a few multi-wavelength campaigns,
during the lifetime of the {\it EGRET/CGRO} detector, that permitted
measurements of time lags or correlations between flares observed in
the $\gamma$-ray band and at longer wavelengths.  Undoubtedly, the
most successful was the one at the beginning of 1996, aimed at 3C 279
(Wehrle et al. 1998). During this campaign, simultaneous coverage was
obtained in practically all frequency bands.

A particularly interesting result of the campaign was the very close
match between the X-ray and $\gamma$-ray light curves during the
Feb-96 outburst event. There was no time lag between the peaks, and
both decayed at the same rate (Lawson, McHardy \&
Marscher~1999). This is not what one would expect, assuming that ERC
dominates the production of radiation down to the X-ray bands, because
the production of X-rays by ERC involves low energy electrons, which
radiate in the adiabatic regime.  As was shown in Figure~\ref{flara}b,
adiabatic flares decay significantly more slowly than radiative
ones. Therefore, the observed $\gamma$--X-ray correlation seems to
exclude the production of X-rays by the ERC process, unless the decay
rate is determined by the deceleration of the shell and/or a change in
its direction of motion.

We applied our ERC model for $\gamma$-ray production during the Feb-96
outburst and found that the main contribution to the X-ray band is
coming from the SSC process. This interpretation of the origin of the
X-rays is consistent with both the time-averaged fit to the spectrum
and the correlation between the $\gamma$-ray and X-ray flares (see
Figure~\ref{3c279} upper panels).  It should be noted, however, that
our model predicts comparable X-ray and $\gamma$-ray flare amplitudes,
while observations show that the $\gamma$-ray flux jumped during the
outburst by a larger factor than the X-ray flux did. But this can be
explained by dilution of the X-ray flare by radiation produced at
larger distances in the jet. It is worth mentioning that even stronger
dilution is required in the optical range, where the outburst was
hardly visible (Wehrle et al. 1998).

In Figure~\ref{3c279}c we also present our X-ray flare profile
superimposed on observations taken during the Feb-96 campaign. Data points
represent $2-10$ keV flux measured by $RXTE$ (Lawson et al. 1999). 
As one can see the ``fit'' is not perfect, especially on its rising side, and 
shows significant deviation from the exponential symmetrical profile used by Lawson et al. (1999). We attribute this discrepancy to two factors. First of all, the model flare in Figure~\ref{3c279}c is produced by a rectangular injection profile. As can be seen from
Figure~\ref{flara}c, triangular injection profiles tend to produce more
symmetrical flares. Secondly, the rise of the X-ray flux before the
$\gamma$-ray flare may indicate the existence of a flare precursor. Such
precursors may be produced by Comptonization of external radiation by
cold electrons necessarily present in inhomogenities before their
collisions and shock formation (Sikora et al., in preparation).

\subsection{Optical--$\gamma$-ray lag: A mirror model?}

Another noticeable multiwavelength event was recorded in PKS 1406-076
(Wagner et al. 1995). In this case, unlike in the Feb-96 flare in
3C279, the $\gamma$-ray flare was clearly accompanied by an optical
one.  Detailed analyses of this event suggest that the peak of the
$\gamma$-ray flare lagged the optical peak by about one day.  If this
is true and this type of lag is confirmed by future observations, it
would have dramatic implications for radiation scenarios in
blazars. Whereas our model predicts an optical--$\gamma$-ray lag (see
panel [d] in Figure~\ref{lags}), it cannot reproduce such a pronounced
lag as claimed for PKS 1406-076 during the Jan-93 outburst.

A model which can be reconciled with the observed lag is the ``mirror"
model, suggested by Ghisellini \& Madau~(1996). According to this
model the synchrotron and Compton flares are produced by the same
source (i.e., the same shock), but at different distances. Synchrotron
flares are produced closer to the center. Radiation from these flares
is scattered/reprocessed by BEL clouds, then Comptonized by the
propagating shock, giving rise to the delayed $\gamma$-ray flares. In
modeling such a scenario, it is necessary to note that the energetics
of nonthermal flares produced by electrons in the radiative regime is
determined by the rate of relativistic electron injection in the jet,
rather than by the intensity of the external radiation field. Also,
the model should take into account the fact that the $\gamma$-ray flux
dominated the radiation output of PKS 1406-076 during the entire
Jan-93 event.  Hence, the $\gamma$-ray flare must be modulated by
$Q(t')$, whereas the synchrotron flare profile depends on both $Q(t')$
and the time dependence of $u_B'/u_{diff}'$, where $u_B'$ is the
energy density of the magnetic field in the propagating source, and
$u_{diff}'$ is the energy density of externally
rescattered/reprocessed synchrotron radiation of the jet.  Our model
results are presented in Figure~\ref{pks1406}.

We would also like to comment on the possible nature of the material
that scatters and reprocesses synchrotron radiation outside the jet.
Can it really consist of clouds producing broad emission lines? Since
the flux of synchrotron radiation near the jet is much larger than the
flux of radiation from the central engine at this distance, the clouds
near the jet are expected to be ionized to a much higher level than
those which are not exposed to jet radiation. The question is then
whether they can produce strong emission lines. Furthermore, there are
some arguments in favor of the BEL region being identified with
accretion disk winds (Emmering, Blandford \& Shlosman 1992; Murray
\& Chiang 1997; Proga, Stone \& Kallman 2000; Nicastro 2000).  If
this is the case, then explaining the optical--$\gamma$-ray lag
through a mirror model would require a different
reprocessor/rescaterrer than BEL clouds.  Alternatively, the increase
of $u_B'/u_{diff}'$ required between the regions producing the
synchrotron and the Compton peaks can be obtained by posting a source
of seed photons in the vicinity of the jet, at a distance beyond the
region where the shock is launched. Such a source could be provided by
a supernova explosion, but the probability of supernovae on parsec
scales in the vicinity of the jet seems to be rather low.

\subsection{On the origin of the MeV break}

In all previous sections it was assumed that the break in the electron
injection function ($\gamma_m$) is located at lower energies than the
break caused by inefficient radiative cooling of electrons with
$\gamma < \gamma_c$.  With this assumption, the characteristic break
of the high energy spectra in the 1--30 MeV range is related to the
break of the electron energy distribution at $\gamma_c$.  As was
discussed by Sikora~(1997), such an interpretation of the MeV break is
consistent with the observed flare time scales. Assuming that the
distance of flare production is of the order $r_{fl} \sim c t_{fl}
\Gamma^2$, one can find that flares lasting $\sim$days are produced at
distances $10^{17}-10^{18}$cm, and that at such distances ERC
radiation is inefficient at $h\nu < 1-30$ MeV (see also Sikora et
al.~1996 for discussion on cooling efficiency in different regions of
the jet).

However, the time resolution of recent high-energy experiments is not
good enough to reject the possibility that the observed flares are
superpositions of several short-lasting flares, which are produced
much closer to the central engine than the $\sim 1$-day variability
time scales suggest. Due to the stronger magnetic and external
radiation fields at smaller distances, the value of $\gamma_c$ would
be lower and the break would be imprinted at $h\nu \ll 1$ MeV. In
order to explain the observed MeV breaks, one would then have to
assume that this break is related to the break in the electron
injection function.

There are two ways to distinguish between these possibilities, by
studying spectral slopes and by studying flare profiles, both below
and above the MeV break. The first approach is based on the fact that
for $\gamma_m < \gamma_c$, the slope of the radiation spectrum
produced by electrons with $\gamma_m \ll \gamma \ll \gamma_c$ should
be harder by $\delta \alpha = 0.5$ than the slope of the radiation
spectrum produced by electrons with $\gamma \gg \gamma_c$.  But in the
case $\gamma_c < \gamma_m$, the slope of the radiation spectrum
produced by electrons with $\gamma_c \ll \gamma \ll \gamma_m$ is
predicted to be $\alpha = 0.5$, independent of the spectral slope
produced at $\nu \gg \nu(\gamma_m)$ (see, e.g., Sari, Piran \& Narayan
1998).  The second approach, proposed in this paper, uses the fact
that adiabatic flares decay much more slowly than radiative ones (see
Figure~\ref{flara}).  Thus, in the case $\gamma_m < \gamma_c$, the
flares observed in OVV/HP quasars at $< $ MeV energies should decay
significantly more slowly than in the case $\gamma_c < \gamma_m$. We
present our predictions for two models, one in which $d\gamma/dt'
=$ const (Figure~\ref{breaka}) and one in which $d\gamma/dt' \propto
1/r^2$ (Figure~\ref{breakb}).  In the latter model the difference
between the two cases is narrowed by the fact that $\gamma_c$ is not
constant.

\section{CONCLUSIONS}

We have studied flare production by thin shells propagating at
relativistic speeds down a jet and applied this model to $\gamma$-ray
quasars.  Our main results can be summarized as follows:

\refitem
-- the sharp nature of flares observed in OVV/HP quasars (see, e.g.,
Mattox et al. 1997; Wagner et al. 1995; Wehrle et al. 1998) seems to
conflict with models in which the flare production time scale is much
longer than the light travel time across the source.  Such models
produce flares which have profiles determined by the injection history
$Q(t')$, and, unless $Q$ is a very sharply peaked function of time,
these flares are too shallow to be consistent with observations;

\refitem 
-- the fast decay of the X-ray flare observed during the Feb-96 event
in 3C 279 suggests that the production of X-rays in this object is
dominated by the SSC process;

\refitem
-- the claimed lag of the $\gamma$-ray flare behind the optical flare
during the Jan-93 outburst in PKS 1406-076 requires a scenario in
which both the energy density of the external radiation field and the
injection rate of electrons increase between the production of the
synchrotron peak and of the ERC peak;

\refitem
-- interpretation of the MeV break in terms of inefficient cooling of
electrons radiating at lower energies predicts a much slower decay of
MeV flares than of GeV flares.

\noindent
The above conclusions can be weakened, or even invalidated, if the
flare decay profiles are determined by deceleration of the shell or a
changing direction of motion. Such a possibility can be tested by
future simultaneous observations of flares at GeV, MeV and keV
energies.  Since kinematically determined flare decays should be
achromatic, finding an example in which X-ray flares decay as quickly
as GeV flares, whereas MeV flares do not, would argue against the role
of kinematics in shaping the flare profiles.

\acknowledgments

This project was partially supported by Polish KBN grant 2P03D 00415,
NSF grants AST-9876887 and AST-9529175, and NASA grant NAG5-6337.
M.S. thanks JILA, University of Colorado, for its hospitality during
the completion of part of this work.

\clearpage

\centerline {\bf FIGURE CAPTIONS}

\figcaption[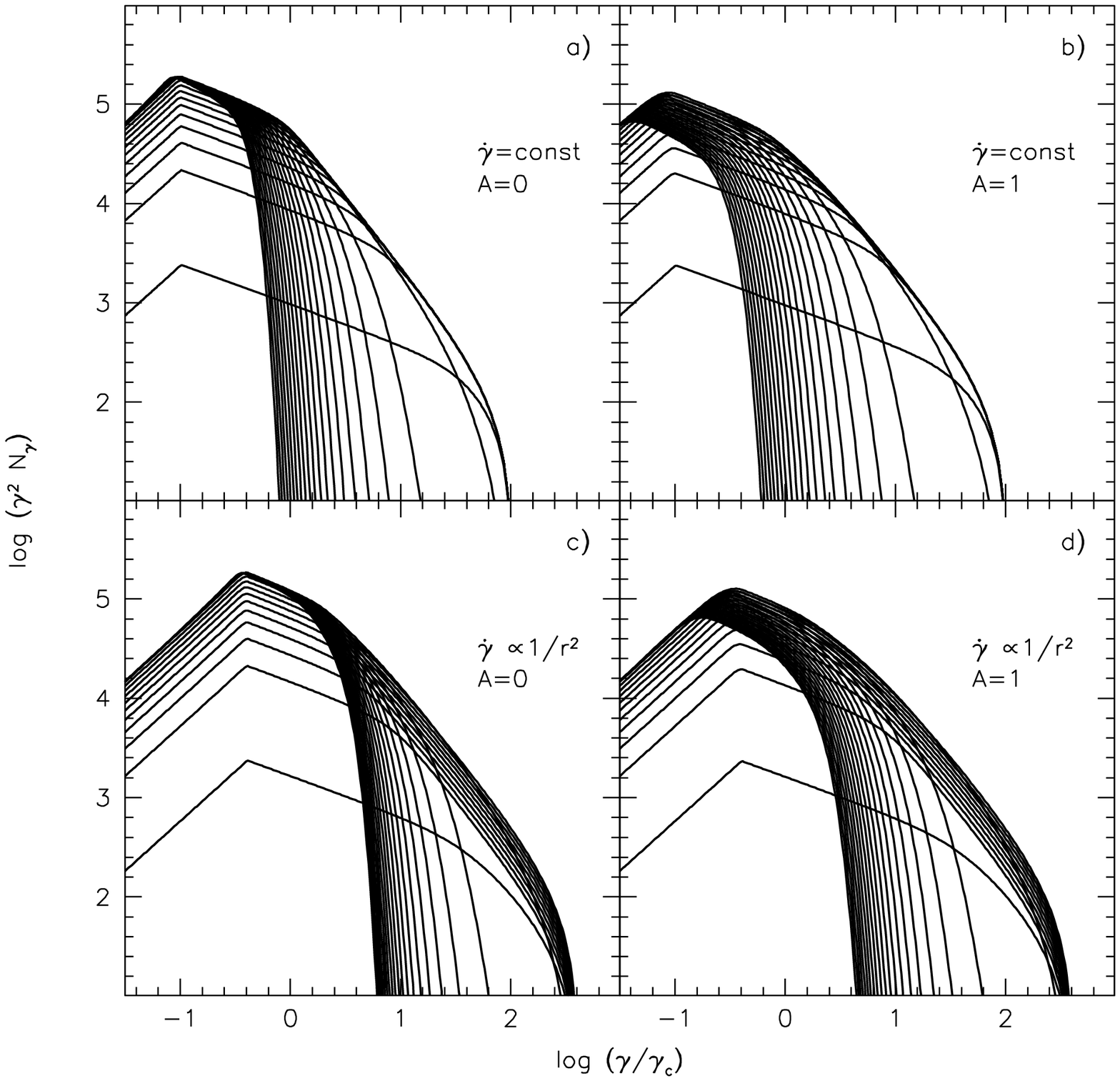]{Evolution of electron energy distribution.
The upper panels are for radiative electron energy loss rates $\propto \gamma^2$ and 
independent of distance [$\dot \gamma (r) \equiv (d\gamma/dt')_{rad} =$
const]; the lower panels are for $\dot \gamma \propto 1/r^2$; the left
panels are for zero adiabatic energy losses, the right panels are for
3D-expansion adiabatic losses. The models are computed assuming
$p=2.4$, $\gamma_m=0.1 \gamma_c$, and $\Delta r_{inj}/r_0
=1$. Evolution is followed from $t/t_0=1$ up to $t/t_0=3$, and the
time step between the presented curves is $t/t_0 =0.1$.
\label{elevol}}

\figcaption[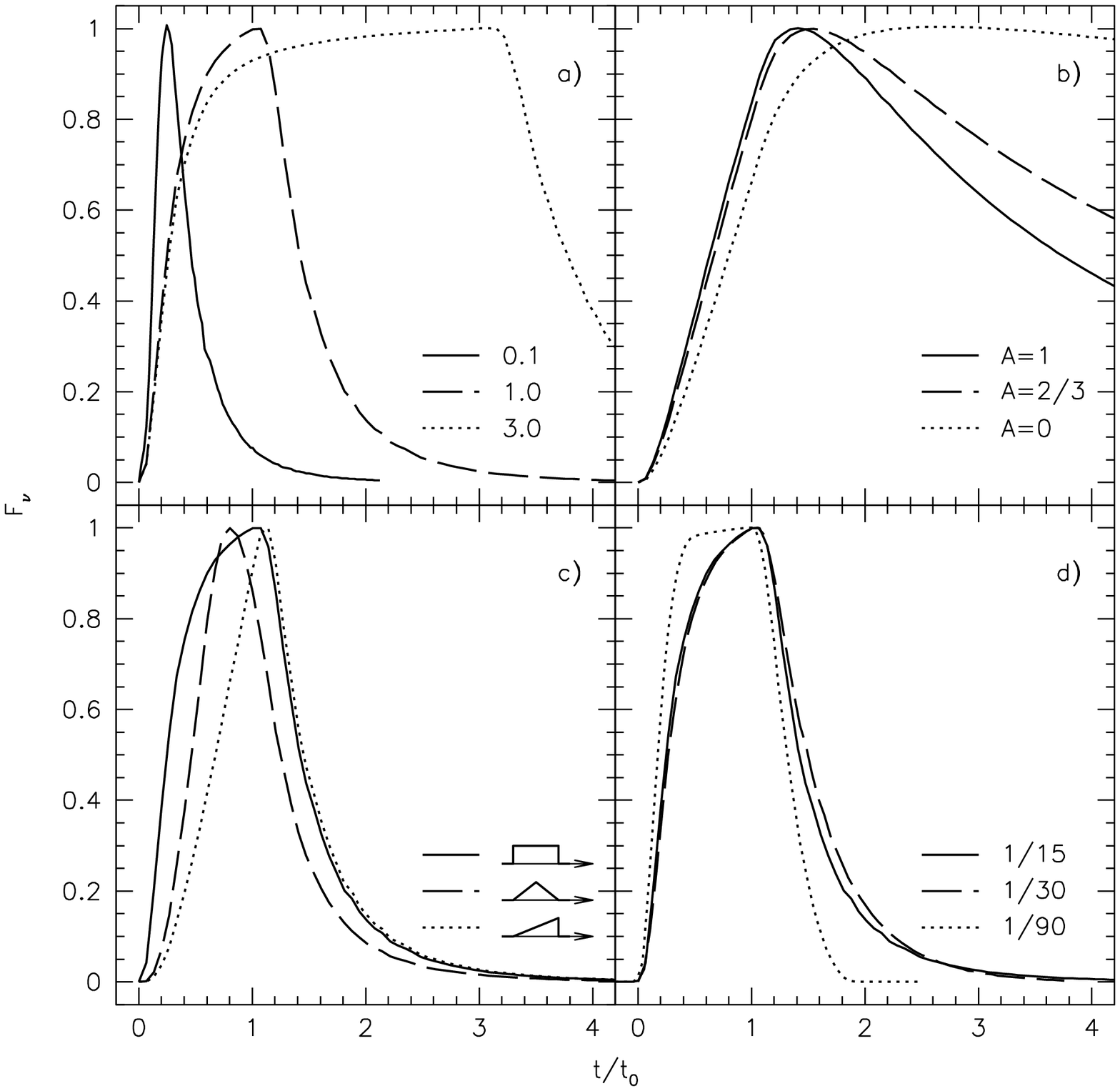]{Flare profiles: (a) for different values of
the injection distance range, $\Delta r_{inj}/r_0$; (b) for different
adiabaticities, $A$; (c) for different electron injection time profiles, $Q(t')$; ; (d) for different jet opening-angles, $\theta_j$.
Models (a), (c), and (d) are computed for $\gamma \gg \gamma_c$
(radiative regime) and models (b) are computed for $\gamma \ll
\gamma_c$ (adiabatic regime).  Unless marked differently, the
parameters used are: $\Delta r_{inj}/r_0=1$; $Q(t')=$ const; $A=2/3$;
$\theta_j = 1/15$.  All models are computed for $\Gamma=15$.
\label{flara}}

\figcaption[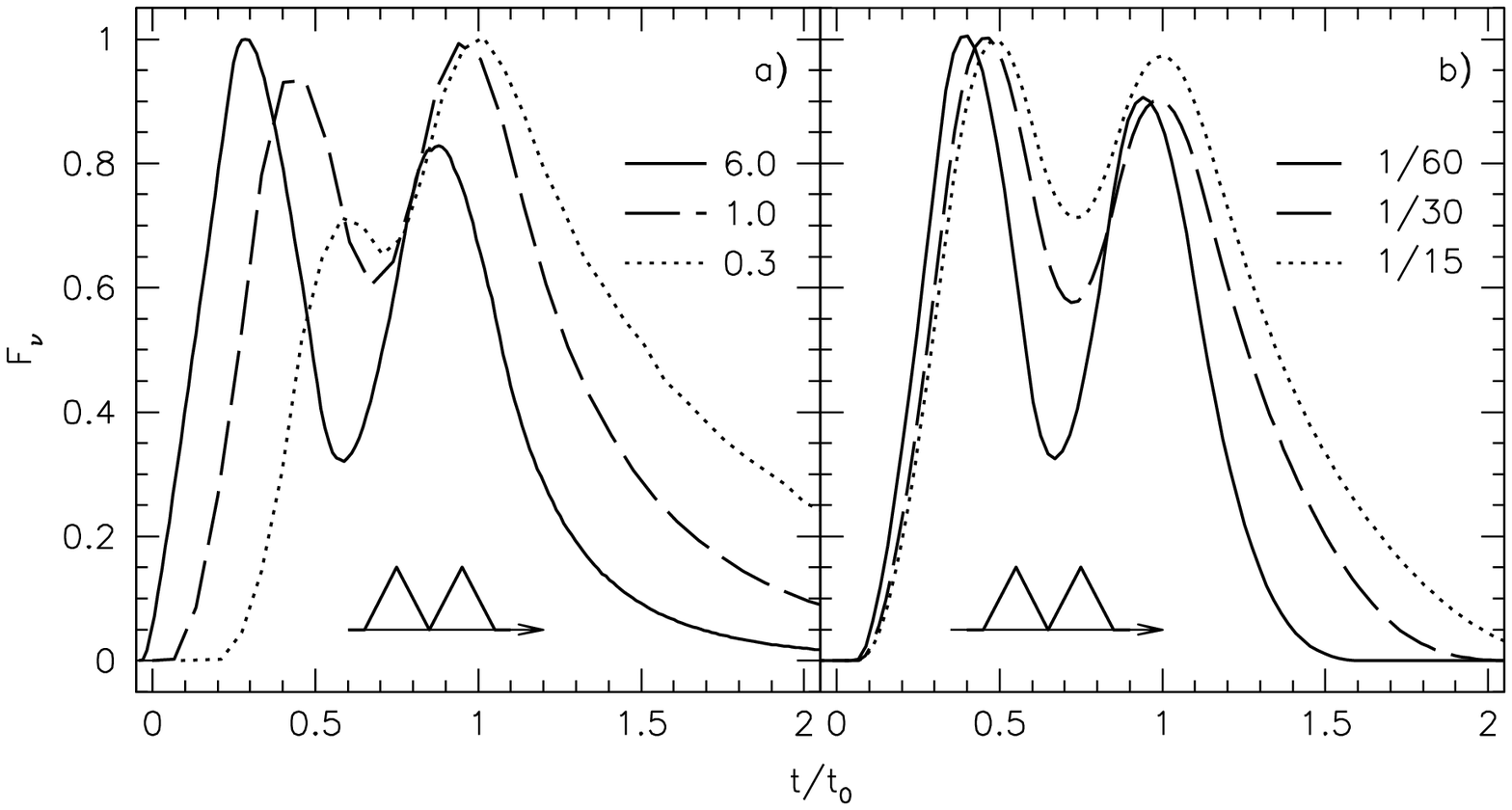]{Flares produced from double-triangle injection
time profile. Panel (a) shows dependence on the injection distance
range $\Delta r_{inj} = 0.3$, $1.0$, $6.0$ per triangle, while panel
(b) is for different observers located at angle $\theta_j = 1/60$,
$1/30$ from the jet axis.
\label{flarb}}

\figcaption[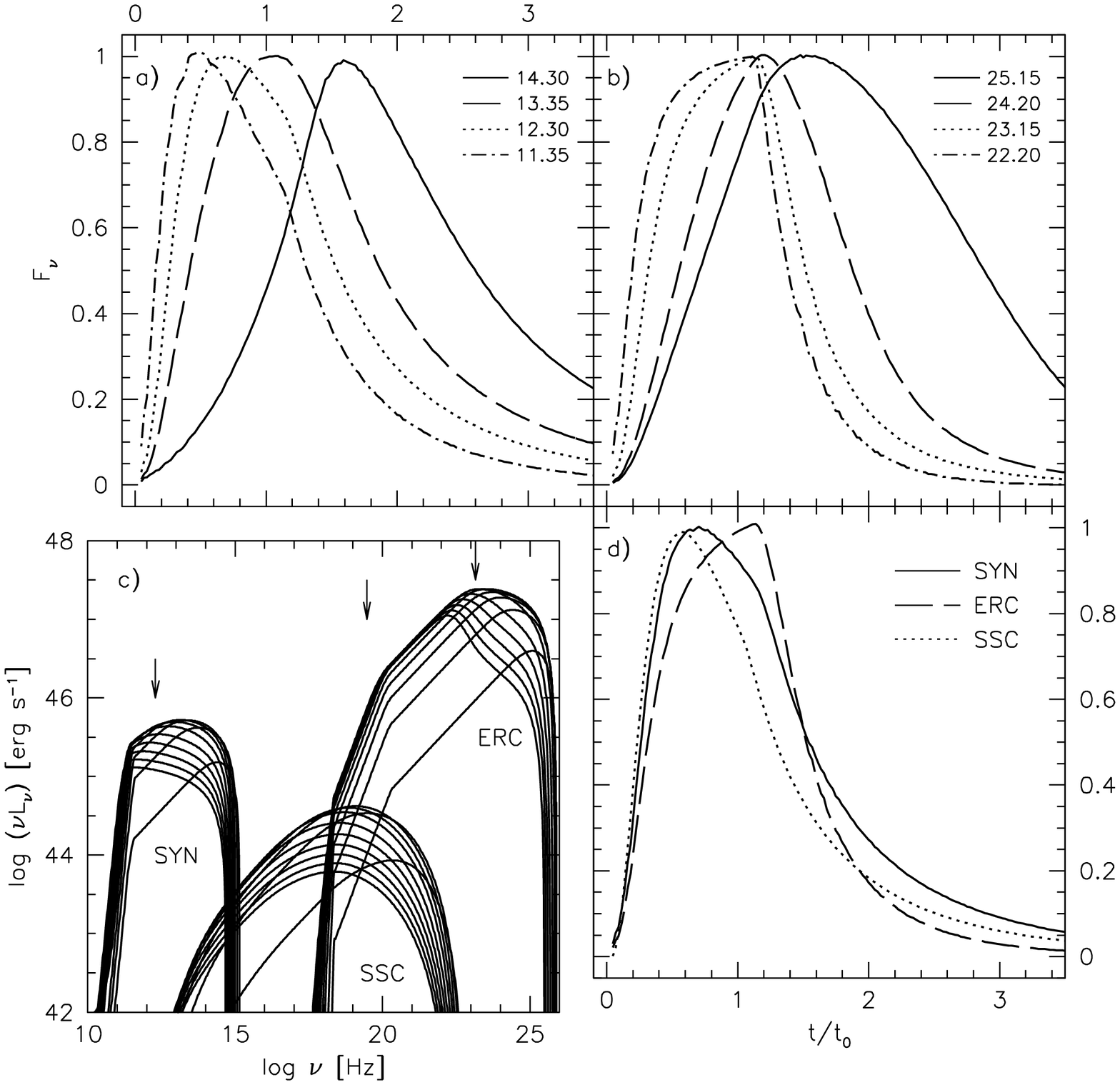]{Flare lags: (a) for different synchrotron frequencies;
(b) for different ERC frequencies; (d) for different radiation
processes, at frequencies as marked on (c). The model has been
computed for the following parameters: bulk Lorentz factor,
$\Gamma=15$; initial injection distance, $r_0 = 6.0 \times 10^{17}$
cm; injection distance range, $\Delta r_{inj}/r_0 =1$; constant
injection rate $Q(t')=$ const, with $\gamma_m = 10$, $\gamma_{max}=6.0
\times 10^3$ and $p=2.2$, and normalization given by $L_e \equiv m_e
c^2 \int Q \gamma ~\ d\gamma = 3.5 \times 10^{43}$ ergs s$^{-1}$;
adiabaticity $A=1$; magnetic field intensity, $B' = 0.4 (r_0/r)$ Gauss;
and energy density of the ambient diffuse radiation field, $u_D=
3.5\times 10^{-4}$ ergs cm$^{-3}$. \label{lags}}
  
\figcaption[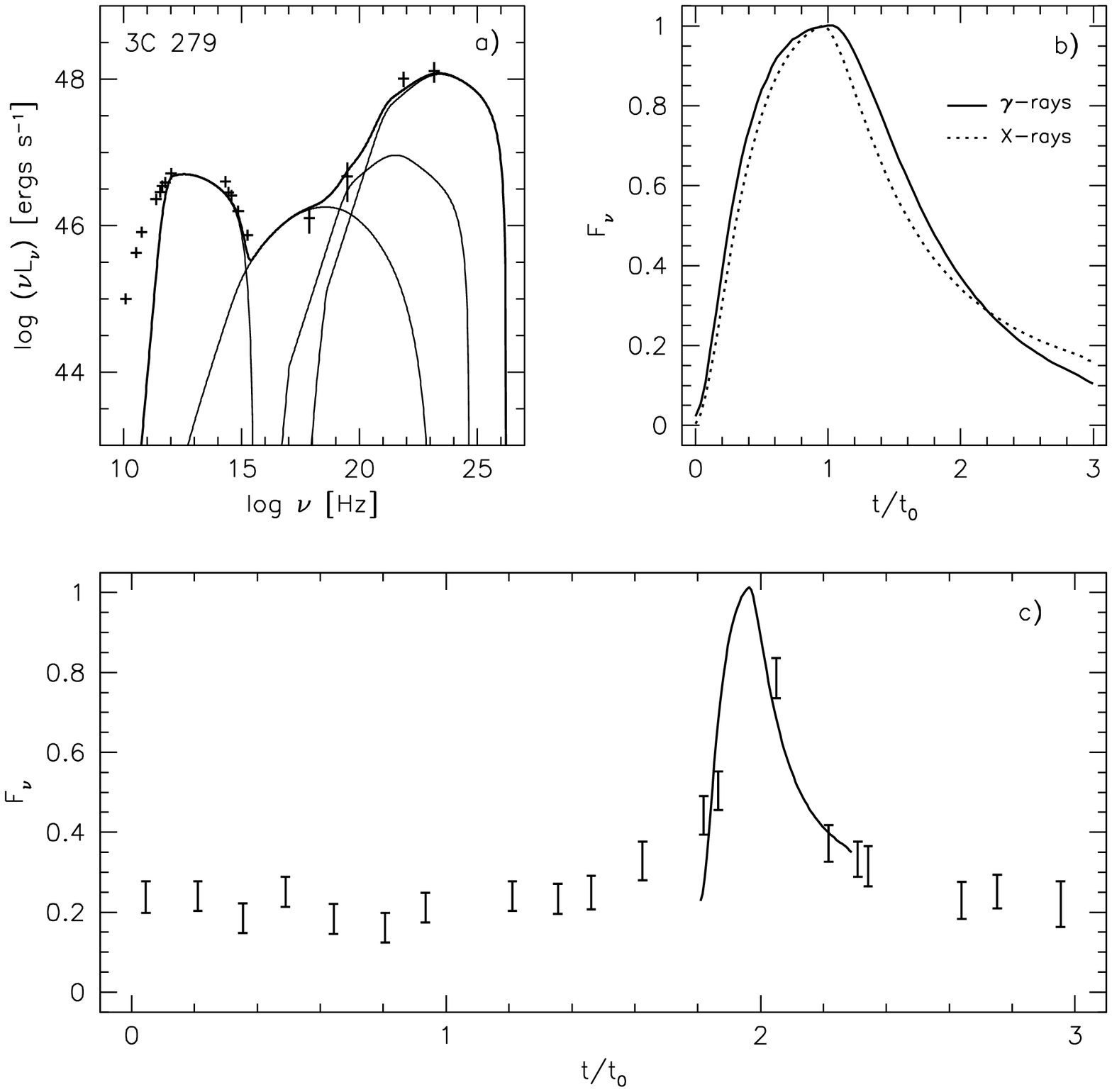]{Modeling the Feb-96 event in 3C 279. 
Panel (a) shows model fit to the time averaged outburst spectrum. All
observational data are simultaneous and taken from Wehrle et
al. (1998). Panel (b): model $\gamma$-ray flare, with the flux
integrated over energies $> 100$ MeV, vs. model X-ray flare, with the
flux taken at 2 keV. Parameters of the model are: $\Gamma=20$; $r_0 =
6.0 \times 10^{17}$ cm; $\Delta r_{inj}/r_0 =1$; $Q(t')=$ const, with
$\gamma_m = 27$, $\gamma_{max}=6.5 \times 10^3$ and $p=2.4$, and $L_e
= 1.6 \times 10^{44}$ ergs s$^{-1}$; $B' = 0.53 (r_0/r)$ Gauss; and
energy densities of the ambient diffuse radiation fields, $u_{BEL}=
4.9 \times 10^{-4} (r_0/r)^2$ ergs cm$^{-3}$ and $u_{IR}= 1.0 \times
10^{-5}$ ergs cm$^{-3}$. Panel (c) presents X-ray flare profile from
panel (b) fitted to $2$-$10$keV $RXTE$ data from the campaign (Lawson,
McHardy \& Marscher~1999).
\label{3c279}}

\figcaption[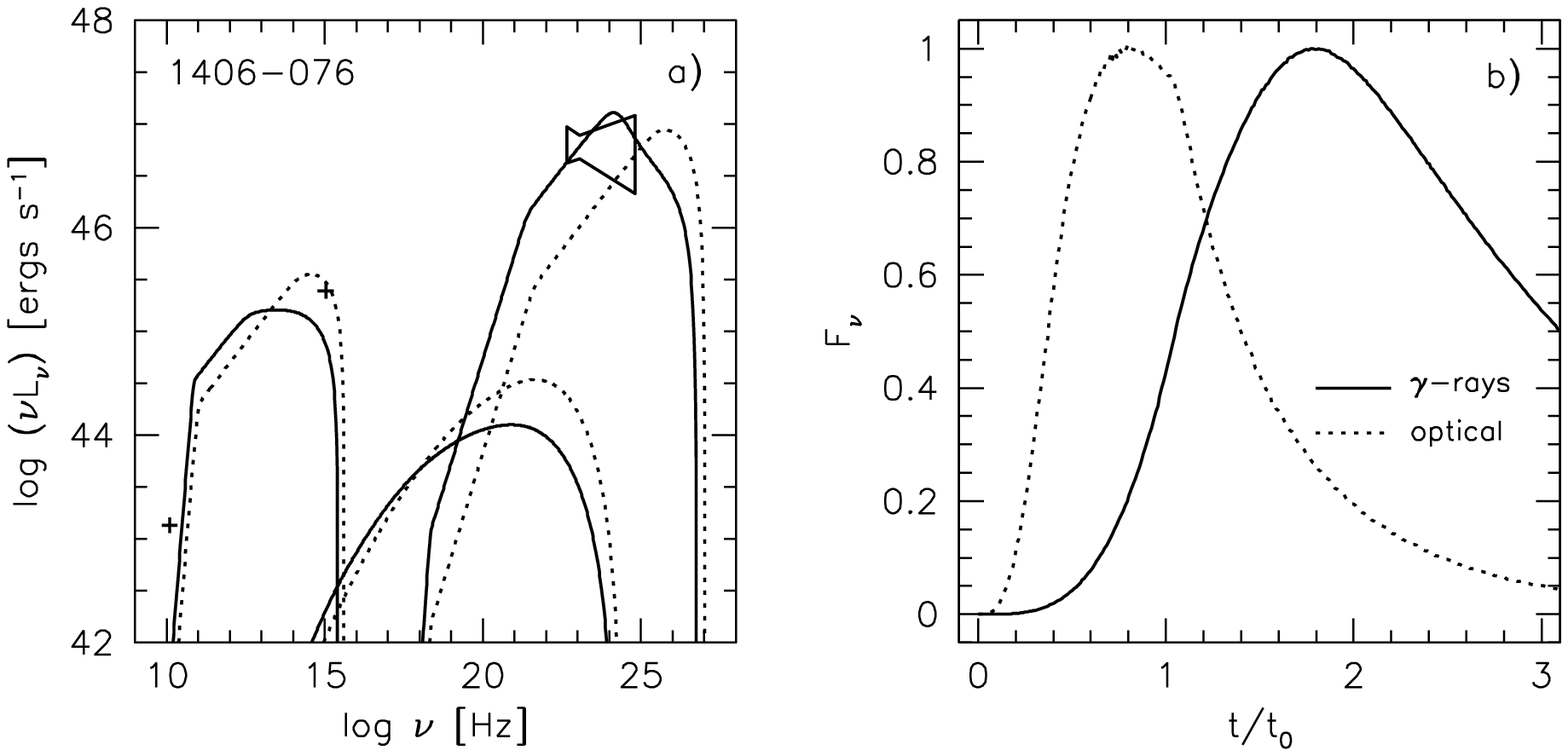]{Modeling  the Jan-93 event in PKS 1406-076.
Left panel: model spectra taken at the moment of the optical flux peak
(dotted line) and at the moment of the $\gamma$-ray flux peak (solid
line).  The optical and $\gamma$-ray data are the fluxes averaged over
the event.  (Radio flux comes from a different epoch.) Right panel:
model $\gamma$-ray flare, with the flux integrated over energies $>
100$ MeV.  vs. model optical flare, taken in the R-band.  The model
parameters are: $\Gamma=15$; $r_0 = 1.2 \times 10^{18}$ cm; $\Delta
r_{inj}/r_0 =1$; $Q(t') \propto (r-r_0)$, with $\gamma_m = 50$,
$\gamma_{max}=2.2 \times 10^4$, $p=2.2$, and $L_e (t=2t_0) = 3.5
\times 10^{43}$ ergs s$^{-1}$; $B' = 0.14 (r_0/r)$ Gauss; and energy
density of the ambient diffuse radiation fields, $u_{IR}= 3.0 \times
10^{-5} (r-r_0)/r_0$ ergs cm$^{-3}$.
\label{pks1406}} 

\figcaption[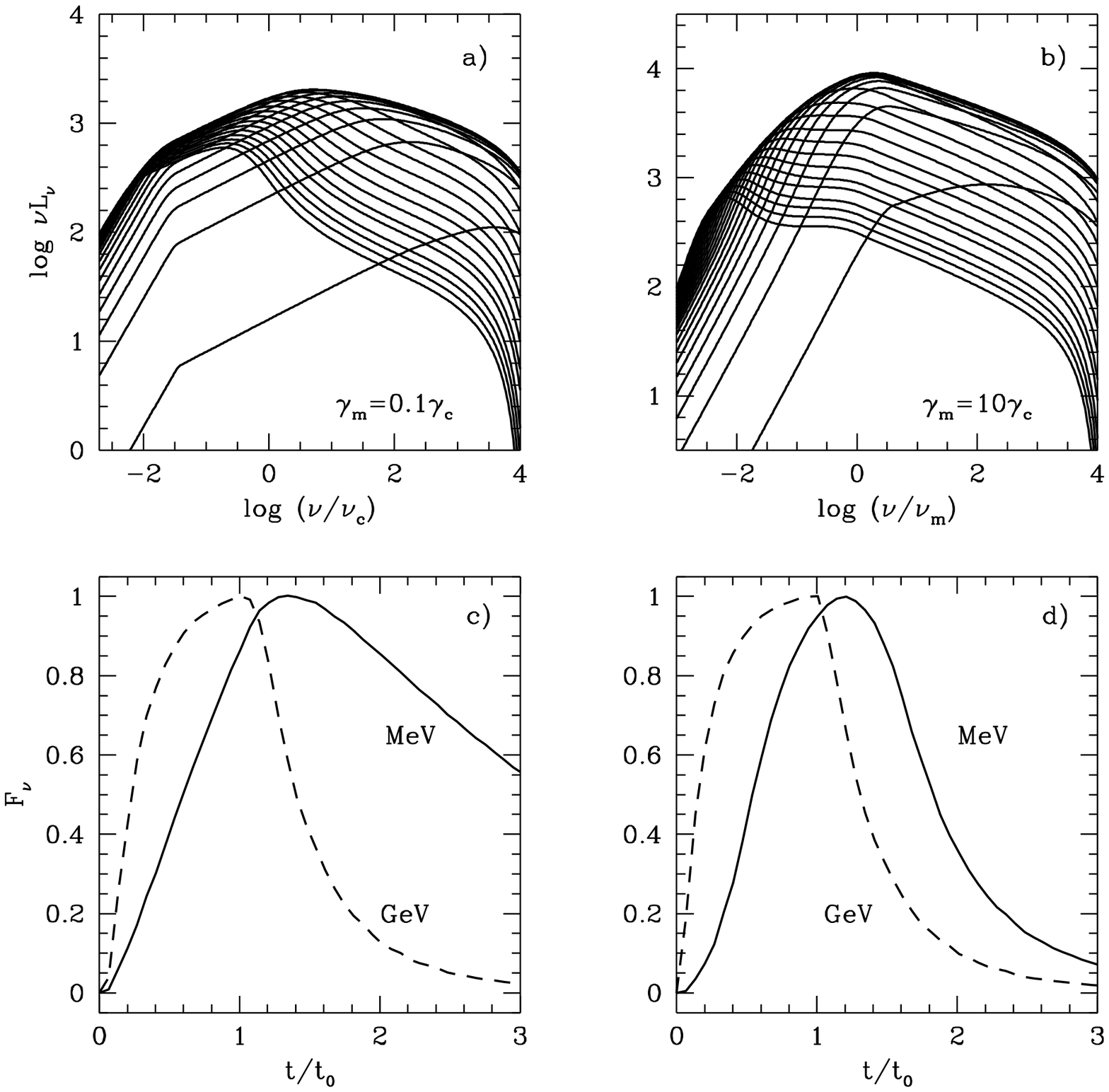]{Variability around the ERC peak.
Left panels are for the peak produced by electrons with energies
$\gamma \sim \gamma_c$, while right panels are for the peak determined
by electrons with $\gamma \sim \gamma_m$. The spectra are shown with a
time step $\delta t/t_0 = 0.14$ and are followed up to $t = 3t_0$.
The flares are shown at frequencies: (c) $\nu = 0.1 \nu_c$ (solid
line) and $\nu = 100 \nu_c$ (dashed line), where $\nu_c = (16/9)
\Gamma^2 \gamma_c^2 \nu_{IR}$; (d) $\nu = 0.1 \nu_m$ (solid line) and
$\nu = 100 \nu_m$ (dashed line), where $\nu_m = (16/9) \Gamma^2
\gamma_m^2 \nu_{IR}$.  $(d\gamma/dt')_{ERC}$ is assumed to be distance
independent.
\label{breaka}}

\figcaption[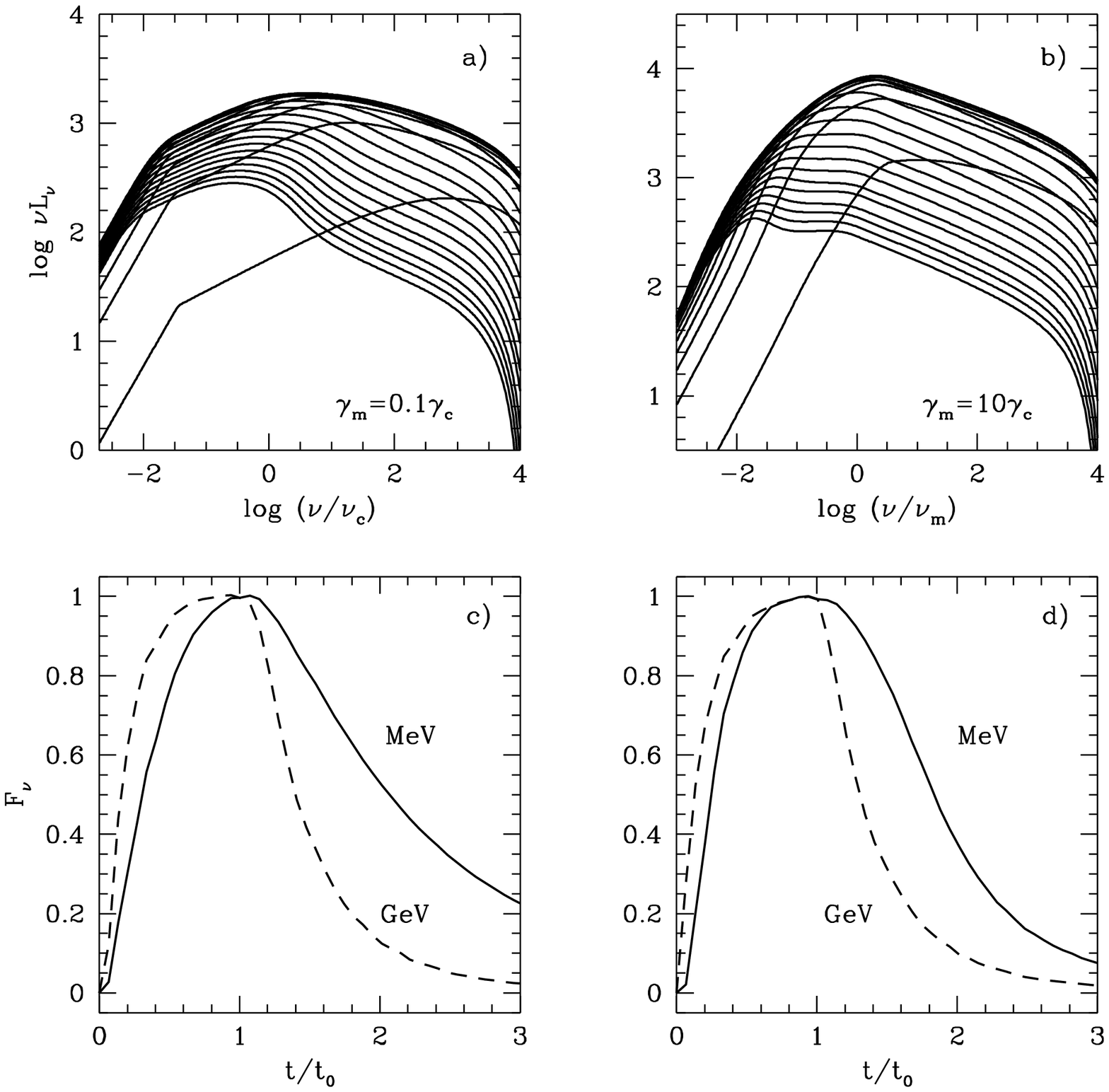]{The same as Fig. 7, but for 
$(d\gamma/dt')_{ERC} \propto 1/r^2$.
\label{breakb}}

\clearpage\centerline{\psfig{file=fig1.ps,height=5.3 in,angle=0}}
\vfill\eject
\centerline{\psfig{file=fig2a.ps,height=5.3 in,angle=0}}
\vfill\eject
\centerline{\psfig{file=fig2b.ps,height=5.3 in,angle=0}}
\vfill\eject
\centerline{\psfig{file=fig3.ps,height=5.3 in,angle=0}}
\vfill\eject
\centerline{\psfig{file=fig4.ps,height=5.3 in,angle=0}}
\vfill\eject
\centerline{\psfig{file=fig5.ps,height=5.3 in,angle=0}}
\vfill\eject
\centerline{\psfig{file=fig6.ps,height=5.3 in,angle=0}}
\vfill\eject
\centerline{\psfig{file=fig7.ps,height=5.3 in,angle=0}}

\end{document}